\documentclass[prl,superscriptaddress,a4paper]{revtex4}
\usepackage{epsfig}
\usepackage{amssymb}
\usepackage{amsmath}
\usepackage{graphicx}
\usepackage{bm}

\begin{document}

\title {Simultaneous cooling and entanglement of mechanical modes of a micromirror in
an optical cavity}

\author{Claudiu Genes, David Vitali
\footnote[3]{To whom correspondence should be addressed (david.vitali@unicam.it)},and Paolo Tombesi }
\address{Dipartimento di Fisica,
Universit\`a di Camerino, I-62032 Camerino, Italy }

\begin{abstract}
Laser cooling of a mechanical mode of a resonator by the radiation pressure of a detuned optical cavity mode has been recently demonstrated by
various groups in different experimental configurations. Here we consider the effect of a second mechanical mode with a close, but different
resonance frequency. We show that the nearby mechanical resonance is simultaneously cooled by the cavity field, provided that the difference
between the two mechanical frequencies is not too small. When this frequency difference becomes smaller than the effective mechanical damping of
the secondary mode, the two cooling processes interfere destructively and cavity cooling is suppressed in the limit of identical mechanical
frequencies. We show that also the entanglement properties of the steady state of the tripartite system crucially depend upon on the difference
between the two mechanical frequencies. If the latter is larger than the effective damping of the second mechanical mode, the state shows fully
tripartite entanglement and each mechanical mode is entangled with the cavity mode. If instead the frequency difference is smaller, the steady
state is a two-mode biseparable state, inseparable only when one splits the cavity mode from the two mechanical modes. In this latter case, the
entanglement of each mechanical mode with the cavity mode is extremely fragile with respect to temperature.
\end{abstract}
\maketitle


\section{Introduction}

Mechanical resonators at the micro- and nano-meter scale are now widely employed in the high-sensitive detection of mass and forces
\cite{roukpt}. Among the applications that have become possible are measurements of forces between individual biomolecules \cite{friedsam},
forces arising from magnetic resonance of single spins \cite{rugarmfm}, and perturbations that arise from mass fluctuations involving single
atoms and molecules \cite{ekinci}. The recent improvements in the nanofabrication techniques suggest that in the near future these devices will
reach the regime in which their sensitivity will be limited by the ultimate quantum limits set by the Heisenberg principle, as first suggested
in the context of the detection of gravitational waves by the pioneering work of Braginsky and coworkers \cite{bragbook}. An important step in
this direction would be the demonstration of cooling such microresonators to their quantum ground state. It would represent a remarkable
signature of the quantum behavior of a macroscopic object, allowing to shed further light onto the quantum-classical boundary \cite{found}.
Recent experiments achieved promising results via cryogenic cooling \cite{schwab}, via back-action cooling, in which the off-resonant operation
of the cavity results in a retarded back action on the mechanical system \cite{gigan06,arcizet06b,vahalacool,mavalvala,harris,schliesser}, or by
cold-damping quantum feedback where the oscillator position is measured through a phase-sensitive detection of the cavity output and the
resulting photocurrent is used for a real-time correction of the dynamics \cite{cohadon99,arcizet06,bouwm,rugar,mavalvala2}. As shown by recent
theoretical results \cite{mancini98,brag,courty,quiescence02,kippenberg07,girvin07,genes08,dantan08}, these cooling scheme are in principle
capable to bring the microresonator down to its ground state. Both mechanisms achieve cooling by increasing the damping of the mechanical
resonator so that it becomes insensitive to thermal noise. It is important to analyze all the possible limitations to ground state cooling, also
because this would open up the experimental realization of a number of genuinely quantum phenomena, such as the possibility of entangling an
acoustic mode to a cavity quantum field \cite{prl07}, or to another mechanical oscillator \cite{mancini02,braunstein03,pinard05,meystre08}, and
even continuous variable quantum information protocols such as quantum teleportation \cite{prltelep}, and entanglement swapping \cite{entswap}.

Here we shall consider only back-action cooling and the prototypal situation of an optical Fabry-Perot cavity with a rigid massive mirror at one
end and a lighter, vibrating mirror at the opposite end. An intense laser beam drives a single, well separated, cavity mode which excites many
internal vibrational modes of the oscillating mirror, via the radiation pressure. All the theoretical treatments of back-action cooling have up
to now focused on a single cavity mode-\emph{single mechanical mode} interaction, which is justified when a bandpass filter in the detection
scheme is used, because coupling between the different vibrational modes is typically negligible.

In this paper we extend the treatment of cooling by including the effect of secondary acoustic modes whose resonance frequency is not far from
that of the mechanical mode of interest, so that they cannot be neglected and filtered out by the detection process. We specialize to the case
of a single additional mode and study its effect on cooling and on the entanglement properties of the steady state of the system. We find that
cooling of the main mode crucially depends upon the difference between the two mechanical resonance frequencies. If this difference is larger
than the effective damping of the secondary mode, cooling is not crucially affected by the presence of the adjacent mode, and the cavity mode is
capable of \textit{simultaneously cooling} the nearby mechanical mode close to its ground state. In this regime the cavity mode is entangled to
each mechanical mode and the stationary state shows fully tripartite entanglement. If instead the two mechanical frequencies are closer than the
effective mechanical damping, then the two cooling processes interfere destructively and each mechanical mode is no more cooled. This
destructive interference affects also the stationary entanglement, which becomes extremely fragile with respect to temperature. The steady state
of the system becomes a two-mode biseparable state, (i.e., of ``class 3'' \cite{giedke}), which means inseparable only when one splits the
cavity mode from the two mechanical modes.

This paper is structured as follows. Section II introduces the physics of the optomechanical interaction inside an optical cavity and the
linearized Langevin equations formalism. In Sec. III, we solve the dynamics of two mechanical modes coupled to the cavity field and compare it
to the case of a single mechanical mode. In Sec. IV we characterize the simultaneous entanglement of the two acoustic modes with the cavity
field and also study the tripartite entanglement of the steady state. Section V concludes the paper.

\section{System dynamics}

We consider an optical Fabry-Perot cavity of length $L$ formed by a rigid
massive mirror at one end and a vibrating micromechanical mirror at the
opposite end, driven by a laser with frequency $\omega_{0}$. We shall refer
from now on to this prototypal situation, even though the analysis could be
easily adapted to other cavity geometries, such as the toroidal silica
microcavities of Ref.~\cite{vahalacool,schliesser}. The laser significantly
drives only a single cavity mode with frequency $\omega_{c}$, from which it is
detuned by $\Delta_{0}=\omega_{c}-\omega_{0}$. The motion of the micro-mirror
can be described by the set of its vibrational normal modes, each with its own
resonance frequency $\omega_{j}$ and damping rate $\gamma_{j}$. The
Hamiltonian of the system is
\begin{equation}
H=\hbar\omega_{c}a^{\dagger}a+\sum_{j}\frac{\hbar\omega_{j}}{2}(p_{j}%
^{2}+q_{j}^{2})+H_{int} +i\hbar E(a^{\dagger}e^{-i\omega_{0}t} -ae^{i\omega
_{0}t}), \label{ham}%
\end{equation}
where the cavity field annihilation operator $a$ satisfies the commutation relation $\left[  a,a^{\dag}\right]  =1$, and the mechanical modes
are described by dimensionless position and momentum operators satisfying $\left[  q_{k},p_{j}\right]  =i\delta_{kj}$. Denoting by $\kappa$ the
cavity decay rate, the parameter $E$ is related to the input power $\mathcal{P}%
_{in}$ by $\left\vert E\right\vert =\sqrt{2\mathcal{P}_{in}\kappa/\hbar
\omega_{0}}$. The single cavity mode description is valid in the adiabatic
limit when all the relevant mechanical frequencies $\omega_{j}$ are much
smaller than the cavity free spectral range $c/2L$, which is typically
satisfied for small cavities. In this limit the scattering of photons by the
mirror motion from the driven mode to the other cavity mode is negligible
\cite{law}. The interaction between the cavity mode and the vibrational modes
is described by $H_{int}$ and it is due to the radiation pressure acting on
the surface $S$ of the vibrating mirror. One has \cite{pinard}
\begin{equation}
\label{radpress}H_{int}=-\int_{S}d^{2} r \vec{P}(\vec{r})\cdot\vec{u}(\vec
{r}),
\end{equation}
where $\vec{P}(\vec{r})$ is the radiation pressure field and
\begin{equation}
\label{displfield}\vec{u}(\vec{r})=\sum_{j}\sqrt{\frac{\hbar}{m_{j}\omega_{j}%
}} q_{j}\vec{u}_{j}(\vec{r})
\end{equation}
is the displacement field of the mirror surface at point $\vec{r}$. This field can be written as a sum over the corresponding (dimensionless)
displacement field of each normal mode, $\vec{u}_{j}(\vec{r})$, which is characterized by an effective mass $m_{j}=\rho\int d^{3}r\left\vert
\vec {u}_{j}(\vec{r})\right\vert ^{2}$ ($\rho$ the mirror mass density). We consider a one-dimensional situation, i.e., we assume that the
driving laser and the cavity are perfectly aligned. In this case, light is sensitive only to mirror surface deformations along the cavity axis,
$u_{x}(\vec{r})$, so that Eq.~(\ref{radpress}) becomes
\begin{equation}
\label{radpress2}H_{int}=-\int_{S}d^{2} r P_{x}(\vec{r}) u_{x}(\vec{r}).
\end{equation}
In general, the radiation pressure due to an optical power $\mathcal{P}$
impinging on a mirror with reflection coefficient $\mathcal{R}$ can be written
as
\begin{equation}
\label{radpress3}P_{x}(\vec{r}) = \frac{2 \mathcal{P}}{c} \mathcal{R}
v_{opt}^{2}(\vec{r}),
\end{equation}
with $v_{opt}(\vec{r})$ denoting the spatial structure of the incident optical
field on the mirror surface. Within the cavity, one can rewrite $2
\mathcal{P}/c=\hbar(\omega_{c}/L)a^{\dagger} a $ and also assume $\mathcal{R}
\simeq1$. One ends up with
\begin{equation}
\label{radpress4}H_{int}=-\hbar\sum_{j} G_{0}^{j}a^{\dagger}aq_{j},
\end{equation}
where the optomechanical couplings are given by
\begin{equation}
G_{0}^{j}=\frac{\omega_{c}c_{j}}{L} \sqrt{\frac{\hbar}{m_{j}\omega_{j}}},
\label{coupls}%
\end{equation}
and
\begin{equation}
c_{j} = \int_{S}d^{2} r v_{opt}^{2}(\vec{r}) (u_{j})_{x}(\vec{r})
\end{equation}
is the overlap at the mirror surface between the cavity mode and the $j$-th mechanical mode. Due to the chosen normalization of
$v_{opt}^{2}(\vec{r})$ and $\vec{u}_{j}(\vec{r})$, the overlaps satisfy the condition $0\leq c_{j} \leq 1$.
Eqs.~(\ref{radpress4})-(\ref{coupls}) show that the radiation pressure directly couples the cavity mode only with the mirror collective
displacement operator $q_{eff}=\sum_{j} G_{0}^{j}q_{j}$. When the detection bandwidth involves only a single, isolated, vibrational normal mode
of the microresonator, the collective coordinate $q_{eff}$ is well approximated by the selected normal mode, and the single harmonic oscillator
description usually adopted is justified. In the more general case, one has to include in the dynamical description of the system all the
vibrational normal modes which contribute to the detected signal.

The unavoidable action of damping and noise onto the dynamics associated with the Hamiltonian of Eq.~(\ref{ham}) is described by adopting the
formalism of quantum Langevin equations \cite{gard,giov} which, in the frame rotating at the laser frequency $\omega_{0}$, are given by
\begin{eqnarray}
\dot{q}_{j}  &  =\omega_{j}p_{j}, \label{QLEnonlinear1}\\
\dot{p}_{j}  &  =-\omega_{j}q_{j}-\gamma_{j}p_{j}+G_{0}^{j}a^{\dag}a+\xi
_{j},\\
\dot{a}  &  =-(\kappa+i\Delta_{0})a+i\sum_{j} G_{0}^{j}aq_{j}+E+\sqrt{2\kappa }a^{in}. \label{QLEnonlinear3}
\end{eqnarray}
The cavity input noise is delta correlated in the time domain $\langle a^{in}(t)a^{in,\dag}(t^{\prime})\rangle=\delta(t-t^{\prime})$, while the
mechanical Brownian stochastic forces with zero mean value $\xi_{j}(t)$ are uncorrelated from each other and have the following, generally
non-Markovian, correlation functions
\begin{equation}
\langle\xi_{k}(t)\xi_{j}(t^{\prime})\rangle=\delta_{kj}\frac{\gamma_{j}}%
{2\pi\omega_{j}}\int d\omega e^{-i\omega(t-t^{\prime})}\omega\left[
\coth\left(  \frac{\hbar\omega}{2k_{B}T}\right)  +1\right]  ,
\end{equation}
with $k_{B}$ the Boltzmann constant and $T$ is the temperature of the
reservoir of the micromechanical mirror. However, the involved mechanical
frequencies are never larger than hundreds of MHz and therefore, as discussed
in \cite{genes08} (see also \cite{benguria}), even for cryogenic temperatures
one can make the approximation
\begin{equation}
\label{thermappr}\frac{\gamma_{j} \omega}{\omega_{j}} \coth\left(  \frac
{\hbar\omega}{2k_{B}T}\right)  \simeq\gamma_{j} \frac{2k_{B} T}{\hbar
\omega_{j}} \simeq\gamma_{j}\left(  2n_{j}+1\right)  ,
\end{equation}
where $n_{j}=\left[  \exp\{\hbar\omega_{j}/k_{B} T\}-1\right]  ^{-1}$ is the
mean thermal phonon number of mode $j$. As a consequence, the Brownian noise
can be safely considered Markovian, that is,
\begin{equation}
\label{browncorre2}\left\langle \xi_{k}(t)\xi_{j}(t^{\prime})\right\rangle
\simeq\delta_{kj}\gamma_{j}\left[  (2n_{j}+1) \delta(t-t^{\prime})+i
\frac{\delta^{\prime}(t-t^{\prime})}{\omega_{j}}\right]  ,
\end{equation}
where $\delta^{\prime}(t-t^{\prime})$ denotes the derivative of the Dirac delta.

Ground state cooling is typically achieved when the radiation pressure coupling is strong. This can be obtained when the intracavity field is
very intense, i.e., for high-finesse cavities and enough driving power. In this limit (and if the system is stable) the system is characterized
by a semiclassical steady state with the cavity mode in a coherent state with amplitude $\alpha_{s}$ ($|\alpha_{s}| \gg1$), and a new
equilibrium position for the vibrational modes, displaced by $q_{s}^{j}$. The parameters $\alpha_{s}$ and $q_{s}^{j}$ are the solutions of the
nonlinear algebraic equations obtained by factorizing Eqs.~(\ref{QLEnonlinear1})-(\ref{QLEnonlinear3}) and setting the time derivatives to zero.
They are given by
\begin{eqnarray}
&  q_{s}^{j}=\frac{G_{0}^{j}|\alpha_{s}|^{2}}{\omega_{j}},\\
&  p_{s}^{j}=0,\\
&  \alpha_{s}=\frac{E}{\kappa+i\Delta_{(N)}},
\end{eqnarray}
where the effective detuning $\Delta_{(N)}$ is obtained from $\Delta_{0}$ by subtracting the frequency shift caused by the steady state
radiation pressure
\label{delta}%
\begin{equation}
\Delta_{(N)}=\Delta_{0}-|\alpha_{s}|^{2}\sum_{j}\frac{[G_{0}^{j}]^{2}}%
{\omega_{j}}.
\end{equation}
Then, we linearize Eqs. (\ref{QLEnonlinear1})-(\ref{QLEnonlinear3}) around the steady state values by writing operators as sums of averages plus
fluctuations: $a=\alpha_{s}+\delta a$, $q_{j}=q_{s}^{j}+\delta q_{j}$ and $p_{j}=p_{s}^{j}+\delta p_{j}$. The nonlinear terms $\delta
a^{\dag}\delta a$ and $\delta a\delta q_{j}$ can be ignored when the fluctuations are much smaller than the mean value, and this is certainly
satisfied when $|\alpha _{s}| \gg1$. One therefore arrives at a system of linearized quantum Langevin equations
\begin{eqnarray}
\delta\dot{q}_{j}  &  =\omega_{j}\delta p_{j}, \label{QLElinear1}\\
\delta\dot{p}_{j}  &  =-\omega_{j}\delta q_{j}-\gamma_{j}\delta p_{j}%
+G_{j}\delta X+\xi_{j},\\
\delta\dot{X}  &  =-\kappa\delta X+\Delta_{(N)}\delta Y+\sqrt{2\kappa}%
X^{in},\\
\delta\dot{Y}  &  =-\kappa\delta Y-\Delta_{(N)}\delta X+\underset{j}{%
{\textstyle\sum} }G_{j}\delta q_{j}+\sqrt{2\kappa}Y^{in}. \label{QLElinear4}
\end{eqnarray}
We have chosen the phase reference of the cavity field so that $\alpha_{s}$ is real and positive, we have defined the field quadratures
$\delta X\equiv(\delta a+\delta a^{\dag})/\sqrt{2}$ and $\delta Y\equiv(\delta a-\delta a^{\dag})/i\sqrt{2}$ and the corresponding Hermitian
input noise
quadratures $X^{in}\equiv(a^{in}+a^{in,\dag})/\sqrt{2}$ and $Y^{in}%
\equiv(a^{in}-a^{in,\dag})/i\sqrt{2}$. We have also defined the effective
optomechanical couplings
\begin{equation}
G_{j}=G_{0}^{j} \alpha_{s}\sqrt{2}=\frac{2\omega_{c}c_j}{L}%
\sqrt{\frac{\mathcal{P}_{in} \kappa}{m_{j} \omega_{j} \omega_{0} \left(
\kappa^{2}+\Delta_{(N)}^{2}\right)  }}. \label{optoc}%
\end{equation}

\section{Simultaneous cooling}

At the steady state, the energy of each mechanical mode can be written in terms of the variances of the corresponding position and momentum
operators,
\begin{equation}
U_{j}=\frac{\hbar\omega_{j}}{2}\left[  \left\langle \delta q_{j}%
^{2}\right\rangle +\left\langle \delta p_{j}^{2}\right\rangle \right].
\end{equation}
In the absence of any cooling mechanism one has $U_{j}=\hbar\omega_{j}%
(n_{j}+1/2) $ and therefore when cooling is present we can write $U_{j}%
=\hbar\omega_{j}(n_{j}^{eff}+1/2)$, where $n_{j}^{eff}$ is the mean effective
excitation number of the $j$ mode, corresponding to an effective mode
temperature $T_{j}^{eff}=\hbar\omega_{j}/[k_{B} \ln(1+1/n_{j}^{eff})]$. If one
defines the vector of fluctuations
\begin{equation}
u(t)=\left(  \delta q_{1}(t),\delta p_{1}(t),...\delta q_{j}(t),\delta
p_{j}(t)...,\delta X(t),\delta Y(t)\right)  ^{\intercal},
\end{equation}
and the vector of noises
\begin{equation}
v(t)=\left(  0,\xi_{1}(t),...0,\xi_{j}(t)...,\sqrt{2\kappa}\delta
X^{in}(t),\sqrt{2\kappa}\delta Y^{in}(t)\right)  ^{\intercal},
\end{equation}
then Eqs.~(\ref{QLElinear1})-(\ref{QLElinear4}) can be written in a compact form as
\begin{equation}
\frac{d}{dt}u(t)=A_{(N)}u(t)+v(t),
\end{equation}
where $A_{(N)}$ is the drift matrix that governs the dynamics of the expectation values. Since the evolution is linear and the noise terms in
Eqs.~(\ref{QLElinear1})-(\ref{QLElinear4}) are zero-mean quantum Gaussian noises, the steady state of the fluctuations is a zero-mean
multipartite Gaussian state fully characterized by its correlation matrix $\mathcal{V}$, whose elements are defined as
\begin{equation}
\mathcal{V}_{lm}=\frac{\left\langle u_{l}\left(  \infty\right)  u_{m}\left(
\infty\right)  +u_{m}\left(  \infty\right)  u_{l}\left(  \infty\right)
\right\rangle }{2}. \label{CM}%
\end{equation}
Using standard techniques \cite{parks}, one can determine the steady state
correlation matrix $\mathcal{V}$ by solving the Lyapunov equation
\begin{equation}
\label{lyap}A_{(N)}\mathcal{V}+\mathcal{V}A_{(N)}^{\intercal}=-D,
\end{equation}
where
\begin{equation}
D=\mathrm{diag}[0,\gamma_{1}\left(  2\bar{n}_{1}+1\right)  ,...0,\gamma
_{j}\left(  2\bar{n}_{j}+1\right)  ,...\kappa,\kappa],
\end{equation}
is the diagonal $(2N+2) \times(2N+2)$ diffusion matrix determined by the noise correlation functions. The stationary variances of the mechanical
modes are given by the corresponding diagonal matrix elements of $\mathcal{V}$.

If only one mechanical mode is considered, the drift matrix assumes the following form

\begin{equation}
A_{(1)}=%
\begin{pmatrix}
0 & \omega_{1} & 0 & 0\\
-\omega_{1} & -\gamma_{1} & G_{1} & 0\\
0 & 0 & -\kappa & \Delta_{(1)}\\
G_{1} & 0 & \Delta_{(1)} & -\kappa
\end{pmatrix}
,
\end{equation}
where $\Delta_{(1)}=\Delta_{0}-G_{1}^{2}/2\omega_{1} $ is the effective single mode detuning. The stationary variances are given by
$\left\langle \delta q_{1} ^{2}\right\rangle = \mathcal{V}_{11}$ and $\left\langle \delta p_{1}^{2}\right\rangle = \mathcal{V}_{22}$ and their
exact expression is given in \cite{genes08}, where they have been obtained by integrating the spectra obtained from the Fourier transform of the
quantum Langevin equations. The two calculations coincide whenever a Markovian treatment of the quantum Brownian noise acting on the mechanical
modes is made, i.e., when the approximation of Eq.~(\ref{thermappr}) is considered.

As shown in \cite{kippenberg07,girvin07,genes08,dantan08}, cooling occurs when $\Delta_{(1)}\simeq\omega_{1}$, i.e., when the laser is
red-detuned with respect to the cavity mode, and the latter is resonant with the AntiStokes sideband of the laser. In fact, the laser light is
scattered by the oscillating mirror into the Stokes and Antistokes sidebands with frequencies $\omega_{0}\pm\omega_{1}$. The generation of an
AntiStokes photon takes away a vibrational phonon and is responsible for cooling, while the generation of a Stokes photon heats the mirror by
producing an extra phonon. If the cavity is resonant with the Antistokes sideband, cooling prevails and one has a positive net laser cooling
rate $\Gamma$ given by the difference of the scattering rates, $\Gamma=A_- - A_+$. Refs.~\cite{kippenberg07,girvin07,genes08,dantan08} also show
that laser cooling is optimized and can approach ground state cooling in the resolved band limit when $\kappa< \omega_{1}$ (actually when
$\kappa\simeq0.2 \omega_{1}$ \cite{kippenberg07,genes08}). Another important condition for ground state cooling is to have large optomechanical
coupling $G_{1}$, which is obtained for large intracavity power. However $G_{1}$ has an upper bound imposed by the stability conditions
\cite{routh}, which in the case of a single mechanical mode and restricting to positive $\Delta_{(1)}$, reduce to the single inequality
\cite{genes08,prl07},
\begin{equation}
\eta^{(1)}=1-\frac{G_{1}^{2}\Delta_{(1)}}{\omega_{1}(\kappa^{2}+\Delta
_{(1)}^{2})}>0. \label{stab1}%
\end{equation}
However, as we have seen in Sec.~II, the optical mode is always coupled to \emph{all} the mechanical modes with a nonzero overlap $c_j$ with the
cavity mode at the mirror surface. Therefore, the actual stability conditions of the system are determined by the coupling with \emph{all} the
$N$ excited mechanical modes, even the unobserved ones. By applying the Routh-Hurwith criterion \cite{routh}, it is possible to see that, if we
restrict to the cooling regime of positive detunings $\Delta_{(N)}$, there is always one nontrivial stability condition only, which is the
direct $N$-mode generalization of Eq.~(\ref{stab1}),
\begin{equation}
\eta^{(N)}=1-\frac{\Delta_{(N)}}{\kappa^{2}+\Delta_{(N)}^{2}}\sum_{j}%
\frac{G_{j}^{2}}{\omega_{j}}>0. \label{stab3}%
\end{equation}
The violation of this condition leads to a bistable behavior, which has been experimentally verified in \cite{dorsel}. We shall assume that this
stability condition is always satisfied from now on.

\begin{figure}[h]
\centerline{\includegraphics[width=1.0\textwidth]{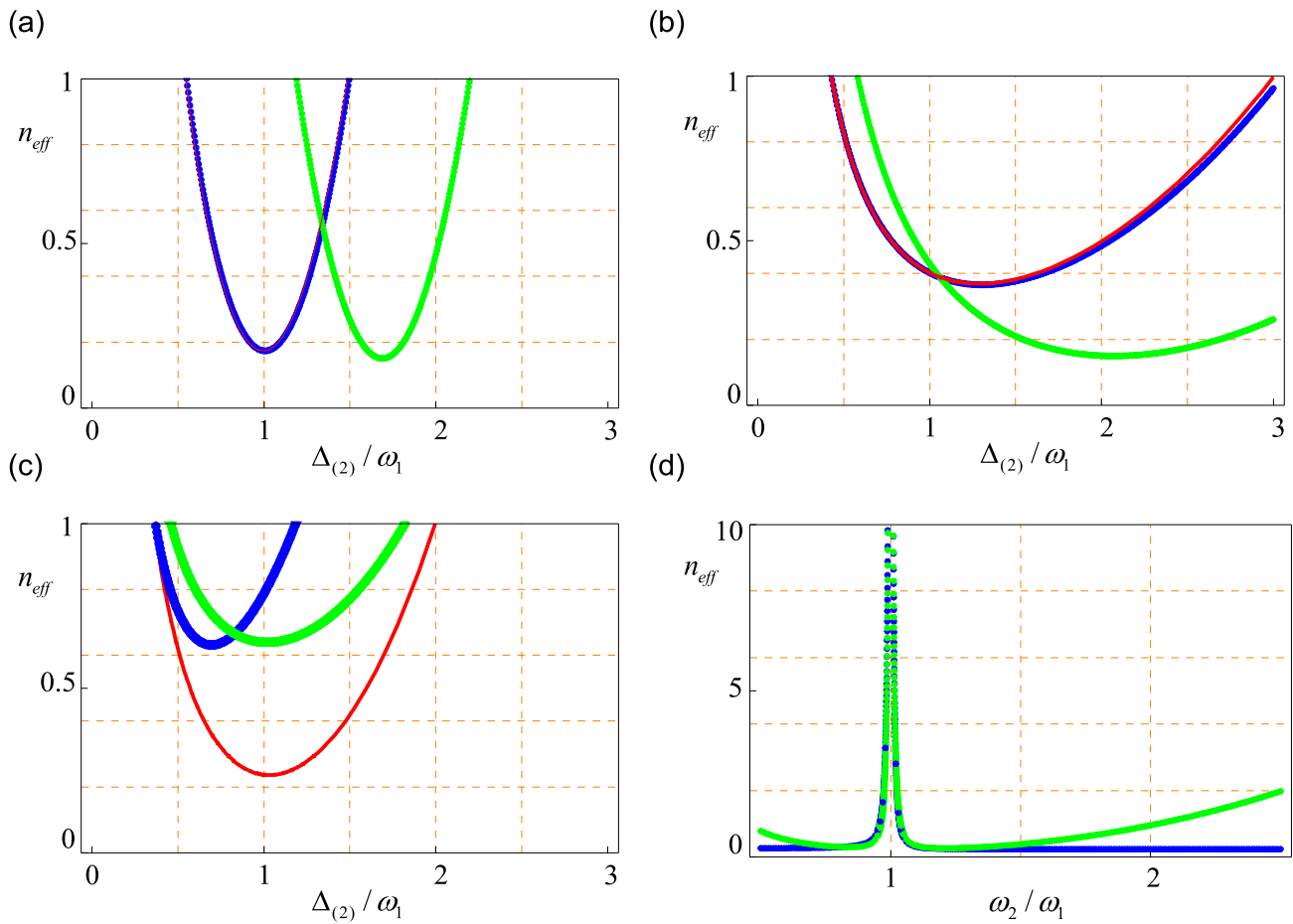}} \caption{a) Simultaneous cooling of two acoustic modes (blue line: main mode,
green line: secondary mode) of a single mirror versus normalized detuning $\Delta_{(2)}/\omega_{1}$. The mechanical parameters are: $\gamma
_{1}/2\pi=\gamma_{2}/2\pi=100$ Hz, $\omega_{1}/2\pi=10$ MHz, $\omega _{2}=1.7\times\omega_{1}$, $m_{1}=m_{2}=250$ ng and $T=0.6$ K, that
corresponds to initial occupancies $n_{1}=1250$ and $n_{2}=735$. The cavity of length $L=1$ mm and finesse $\mathcal{F}=1.5\times10^{5}$ is
driven by a laser of wavelength $\lambda_{0}=1064$ nm and power $P_{0}=30$ mW (at $\Delta_{(2)}=\omega_{1}$), which gives $\kappa\simeq
G_{1}\simeq\omega _{1}/5\,$. A red line is plotted but not visible owing to its almost complete overlap with the blue line, that portraits the
behavior of the main mode cooling in the absence of the secondary mode. The final effective temperatures achieved are $T_{1}^{eff}\simeq0.24$ mK
and $T_{2}^{eff}\simeq0.42$ mK. At the intersection of the cooling curves the effective temperatures are $T_{1}^{eff}\simeq0.46$ mK and
$T_{2}^{eff}\simeq0.79$ mK b) Simultaneous cooling for a smaller cavity finesse $\mathcal{F}=3\times10^{4}$ (corresponding to
$\kappa\simeq\omega_{1}$) and $P=100$ mW (that gives$\ G_{1}=0.6\times\omega_{1}$). The main mode is chosen as before and its independent
cooling curve is shown in red (again the two curves for $n_{eff}$ overlaps almost everywhere), while the secondary mode is fixed at frequency
$\omega_{2}=2\omega_{1}$. c) Simultaneous cooling for closely spaced modes $\omega_{2}=0.95\times\omega_{1}$ and parameters
$\kappa\simeq\omega_{1}/2$, $G_{1}\simeq0.3\times\omega_{1}$. The cooling curve of the main mode (blue line) in the presence of the secondary
mode (green line) is quite different from the independent cooling curve (red line). d) For the same parameters as in (c) the occupancy of the
both modes at $\Delta_{(2)}=\omega_{1}$ is shown versus $\omega_{2}/\omega_{1}$. An optimal occupancy of $0.22$ is reached for well separated
frequencies but it is strongly disturbed around $\omega_{2}/\omega_{1}\simeq 1$. When the two frequencies are equal cooling is practically
absent.} \label{cooling}
\end{figure}

In the case of two mechanical modes, the drift matrix is%

\begin{equation}
A_{(2)}=%
\begin{pmatrix}
0 & \omega_{1} & 0 & 0 & 0 & 0\\
-\omega_{1} & -\gamma_{1} & 0 & 0 & G_{1} & 0\\
0 & 0 & 0 & \omega_{2} & 0 & 0\\
0 & 0 & -\omega_{2} & -\gamma_{2} & G_{2} & 0\\
0 & 0 & 0 & 0 & -\kappa & \Delta_{(2)}\\
G_{1} & 0 & G_{2} & 0 & \Delta_{(2)} & -\kappa
\end{pmatrix}
.
\end{equation}
We have exactly solved the Lyapunov equation (\ref{lyap}) and analyzed the stationary position and momentum variances of the two mechanical
modes in a parameter regime close to that of optimal cooling for a single mode, in order to see if and how the secondary mechanical mode affects
ground state cooling. The results are shown in Fig.~\ref{cooling}, where the effective phonon number $n_{eff}$ of the main mode (blue line) and
of the secondary mode (green line) are plotted and compared to the single mode cooling case (red line). We have considered experimentally
feasible parameters, (see caption) i.e., mechanical quality factors of the order of $10^{5}$ and resonance frequency of the main mode
$\omega_{1}/2\pi=10$ MHz.

We find two different situations, depending upon the value of the difference between the two mechanical frequencies, $\delta
\omega_{21}=\omega_2-\omega_1$. Figs.~\ref{cooling}(a) and \ref{cooling}(b) refer to the case when the two frequencies are well distinct,
$\omega_{2}=1.7\omega_{1}$ in (a) and $\omega_{2}=2\omega_{1}$ in (b), and they plot $n_{eff}$ versus the effective cavity detuning $\Delta
_{(2)}=\Delta_{0}-G_{1}^{2}/2\omega_{1}-G_{2}^{2}/2\omega_{2} =\Delta _{(1)}-G_{2}^{2}/2\omega_{2}$. Fig.~\ref{cooling}(a) refers to the good
cavity limit, $\kappa \simeq 0.2 \omega_1$, corresponding to a cavity finesse $\mathcal{F}=1.5\times10^{5}$, while Fig.~\ref{cooling}(b) refers
to a larger cavity bandwidth, $\kappa \simeq \omega_1$, corresponding to a finesse $\mathcal{F}=3\times10^{4}$. We consider a reservoir
temperature $T=0.6$ K, oscillators with mass $m=250$ ng and a cavity length $L=1$ mm. The results show that, when the two mechanical modes are
well separated ($\delta \omega_{21} \simeq \omega_1$), the nearby mode does not disturb the cooling of the mechanical mode of interest, as
witnessed by the perfect overlap of the blue and red curves, both in (a) and in (b). Even better, the secondary mode is \emph{simultaneously}
cooled close to its ground state (green curve). The comparison between Figs.~\ref{cooling}(a) and \ref{cooling}(b) shows that simultaneous
cooling is influenced by the value of the cavity bandwidth $\kappa$. In fact, the interval for the detuning $\Delta_{(2)}$ within which one has
a significantly low value of $n_{eff}$ is given by $\omega_j-\kappa \lesssim \Delta_{(2)} \lesssim \omega_j+\kappa$, as it can be seen from the
width of the peak of the net laser cooling rate for mode $j$ \cite{kippenberg07,girvin07,genes08,dantan08},
\begin{equation} \Gamma_j =\frac{2G_j^{2}\Delta_{(2)} \omega _{j}\kappa }{\left[ \kappa
^{2}+(\omega_j -\Delta_{(2)} )^{2}\right] \left[ \kappa ^{2}+(\omega_j +\Delta_{(2)} )^{2}\right] } \label{dampeff}
\end{equation}
as a function of $\Delta_{(2)}$. Therefore, if the difference between the two mechanical frequencies is larger than $\kappa$, the two modes are
optimally cooled at two well distinct values of $\Delta_{(2)}$ and one can efficiently cool both modes only by fixing the detuning within a very
narrow interval halfway between the two mechanical resonances, $\Delta_{(2)}\simeq \left(\omega_1+\omega_2\right)/2$ (see
Fig.~\ref{cooling}(a)). However, due to the small value of $\kappa$, the achievable value of $n_{eff}$ at this intermediate detuning is
appreciably larger ($n_{eff}\simeq 0.5$) than the optimal one achievable if one wanted to cool only one mode ($n_{eff}\simeq 0.15$). Instead,
for a larger cavity bandwidth, $\kappa \simeq \delta \omega_{21}$, a good simultaneous cooling of both modes is achievable in a significantly
wider interval of detunings $\Delta_{(2)}$ (see Fig.~\ref{cooling}(b)).

Figs.~\ref{cooling}(c) and \ref{cooling}(d) show that the situation is very different when the two mechanical resonances get very close.
Fig.~\ref{cooling}(c) refer to $\omega_{2}=0.95\times\omega_{1}$, and one can see that in this case the occupancies $n_{eff}$ of the two modes
are appreciably higher than the one corresponding to a single isolated mode (red line). This means that when $\delta \omega_{21}$ is small
enough, the two cooling processes tend to interfere destructively. This is clearly confirmed by Fig.~\ref{cooling}(d), where $n_{eff}$ at the
fixed optimal detuning $\Delta_{(2)}=\omega_1$ is plotted versus the ratio $\omega_2/\omega_1$: an occupancy of $n_{eff}=0.22$ is reached for
well separated frequencies but both modes are practically uncooled in a small interval around $\omega_{2}/\omega_{1} \simeq 1$.

\begin{figure}[h]
\centerline{\includegraphics[width=1.0\textwidth]{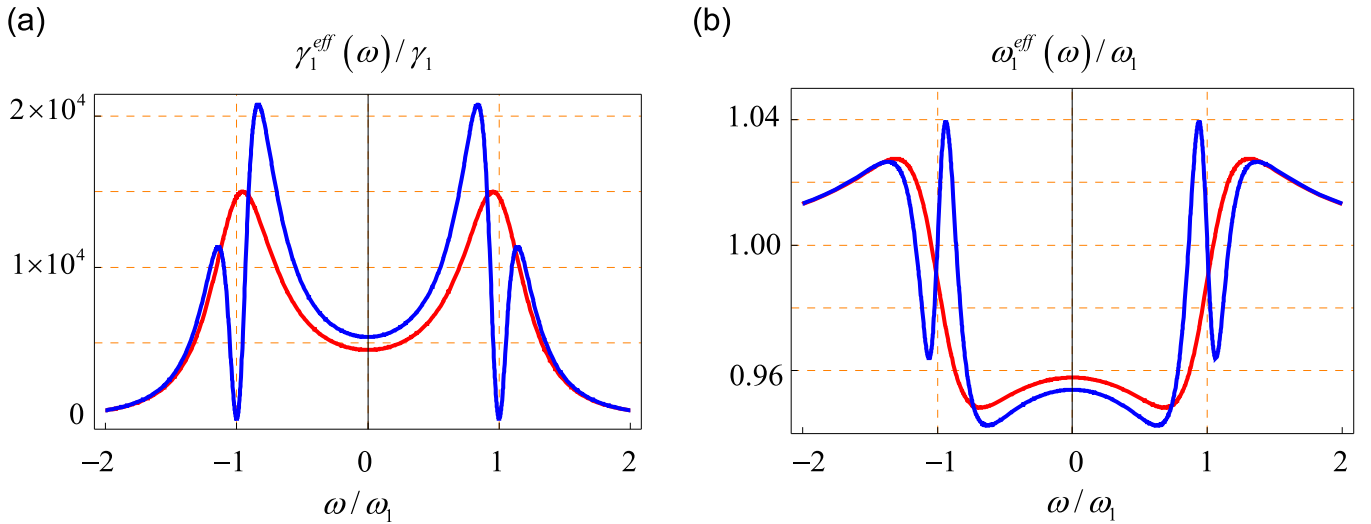}} \caption{Plot of normalized effective mechanical damping rate (a) and
mechanical frequency (b) of the main mode in the absence (red line) and presence (blue line) of a secondary mode as a function of
$\omega/\omega_{1}$. The parameters are $\gamma_{1}/2\pi=\gamma_{2}/2\pi=100$ Hz, $\omega_{1}/2\pi=10$ MHz, $\omega_{2}=\omega_{1}$,
$m_{1}=m_{2}=250$ ng, $T=0.6$ K, $L=0.5$ mm, $\kappa=0.3\times\omega_{1}$ and $G_{1}=\omega_{1}/5$. The quantities are plotted in the optimal
cooling regime where $\Delta
_{(1)}=\omega_{1}$ for the independent cooling case and $\Delta_{(2)}%
=\omega_{1}$ for the simultaneous cooling case. At $\omega=\omega_{1}$, the effect of the radiation pressure on the main mode is suppressed
owing to the coupling to the secondary mode.}
\label{susc}
\end{figure}

This fact, at first sight unexpected, can be explained in terms of classical destructive interference between two resonant oscillators. A first
explanation can be obtained by looking at the mechanical susceptibility of the main oscillator in the presence of the second mode,
$\chi_{1}^{tm}\left( \omega\right)$. It can be derived by Fourier-transforming the quantum Langevin equations, and it is given by
\begin{equation}
\left[  \chi_{1}^{tm}\left(  \omega\right)  \right]  ^{-1}=\left[  \chi _{1}\left(  \omega\right)  \right]  ^{-1}-\chi_{2}\left(  \omega\right)
z^{2}\left(  \omega\right)  G_{1}^{2}G_{2}^{2}, \label{chitm}
\end{equation}
where
\begin{equation}
z\left(  \omega\right)=\frac{\Delta}{(\kappa-i\omega)^{2}+\Delta^{2}},
\end{equation}
\begin{equation}
\left[  \chi_{i}\left(  \omega\right)  \right]  ^{-1}  =\left[  \chi_{0} ^{1}\left(  \omega\right)  \right]  ^{-1}-z\left(  \omega\right)
G_{i}^{2},\;\;\mathrm{i=1,2} \label{chit0}
\end{equation}
is the susceptibility of mode $i$ modified by the radiation pressure of the cavity field, in the absence of the other mode, and
\begin{equation}
\left[  \chi_{0}^{i}\left(  \omega\right)  \right]  ^{-1}=\frac{1}{\omega_{i} }\left[  \left(  \omega_{i}^{2}-\omega^{2}\right)
-i\omega\gamma_{i}\right]
\end{equation}
is the bare susceptibility of the isolated microresonator. One can get an intuitive idea of the response of the mechanical mode of interest by
rewriting $\chi_{1}^{tm}\left( \omega\right)$ as the susceptibility of an harmonic oscillator with frequency-dependent resonance frequency $
\omega_{i}^{eff}(\omega)$ and damping $\gamma_{i}^{eff}(\omega)$,
\begin{equation}
\left[  \chi_{1}^{tm}\left(  \omega\right)  \right]  ^{-1}=\frac{1}{\omega_{i}}\left[  \left(  \omega_{i}^{eff}(\omega)^2-\omega^{2}\right)
-i\omega\gamma_{i}^{eff}(\omega)\right].
\end{equation}
These two functions in the case of mode 1 are plotted in Fig.~\ref{susc} for the case of identical mechanical frequencies, $\omega_1=\omega_2$.
In this case both the effective damping and the effective frequency are strongly modified by the presence of the second mode (blue curve). The
modification of the mechanical frequency due to radiation pressure is the so-called ``optical spring effect'', which may lead to significant
frequency shifts in the case of low-frequency oscillators \cite{mavalvala}, but does not have significant effects in the case of higher
frequencies, such as those of Refs.~\cite{gigan06,arcizet06b,vahalacool}, and assumed here (see Fig.~\ref{susc}(b)). What is relevant is the
modification of the effective damping that, in the presence of the second mode, quickly drops to a very small value in a narrow interval around
resonance. Since cooling is signalled by an increased mechanical damping, this drop is just a manifestation of the suppression of cooling taking
place when the two mechanical modes are resonant (see Fig.~\ref{susc}(a)). This behavior is well described by the analytic expression of the
effective frequency-dependent damping $\gamma_{i}^{eff}(\omega)$. The latter assumes a simple and transparent form when the susceptibility of
mode $i$ in the absence of the other mode can be taken as that of a usual resonator with an unmodified resonance frequency and a
frequency-independent damping rate given by the net laser cooling rate $\Gamma_i$,
\begin{equation}
\left[  \chi_{i}\left(  \omega\right)  \right]  ^{-1}=\frac{1}{\omega_{i}%
}\left[  \left(  \omega_{i}^{2}-\omega^{2}\right)  -i\omega\Gamma_{i}\right]. \label{assum}
\end{equation}
In fact, in this case, inserting Eq.~(\ref{assum}) into Eq.~(\ref{chitm}), one arrives at
\begin{equation}
\gamma_{i}^{eff}(\omega)  \simeq\gamma_{1}+\Gamma_{1}\frac{\left[  \left(  \omega _{2}^{2}-\omega^{2}\right)
^{2}+\omega^{2}\gamma_{2}\Gamma_{2}\right] }{\left(  \omega_{2}^{2}-\omega^{2}\right)  ^{2}+\omega^{2}\Gamma_{2}^{2}}\simeq\left(
\gamma_{1}+\Gamma_{1}\right)  -\Gamma_{1}\frac{\omega ^{2}\Gamma_{2}^{2}}{\left(  \omega_{2}^{2}-\omega^{2}\right)  ^{2}+\omega
^{2}\Gamma_{2}^{2}} . \label{gammaeffbel}
\end{equation}
Since the effective resonance frequency is not altered, the effective damping of mode $1$ in the presence of the second mode is essentially
determined by $\gamma_{i}^{eff}(\omega_1)$, i.e., Eq.~(\ref{gammaeffbel}) evaluated at resonance. Therefore one has that when
$\omega_1=\omega_2$, $\gamma_{i}^{eff}(\omega_1)  \simeq \gamma_1+\Gamma_1(\gamma_2/\Gamma_2) \simeq \gamma_1$ (since typically $\gamma_2\ll
\Gamma_2$), implying that at resonance the second order scattering processes mediating the interaction between the two mechanical modes
completely suppress cooling. This process is a classical destructive interference phenomenon, similar to the classical analogue of
electromagnetically induced transparency realized with two coupled oscillators in Ref.~\cite{paulo}. Another important information provided by
Eq.~(\ref{gammaeffbel}) is that this destructive interference starts to affect cooling when $\delta\omega_{21} < \Gamma_2$, that is,
Eq.~(\ref{gammaeffbel}) shows that the ``bandwidth'' within which one has suppression of cooling is given by the effective mechanical damping,
modified by the radiation pressure of the cavity, $\Gamma_2$.

One can see also view the suppression of cooling taking place when the two mechanical modes are resonant in a different way, starting from the
interaction Hamiltonian of Eq.~(\ref{radpress4}), which shows that the cavity mode directly interacts only with the collective coordinate
$\sum_j G_0^j q_j$. This suggests that, in the case of two mechanical modes, it is useful to pass to the effective ``center-of-mass'' and
``relative'' coordinates
\begin{eqnarray}
q_{cm}&=&\frac{G_0^1 q_1+G_0^2 q_2}{[G_0^1]^2+[G_0^2]^2},\hspace{0.5cm}p_{cm}=\frac{G_0^1 p_1+G_0^2 p_2}{[G_0^1]^2+[G_0^2]^2}, \label{ceadim}\\
q_{r}&=&\frac{G_0^1 q_2-G_0^2 q_1}{[G_0^1]^2+[G_0^2]^2},\hspace{0.5cm}p_{r}=\frac{G_0^1 p_2-G_0^2 p_1}{[G_0^1]^2+[G_0^2]^2}.
\end{eqnarray}
With the new coordinates, the free Hamiltonian of the two mechanical modes becomes
\begin{equation}
H_{mech}=\frac{\hbar \omega_{cm}}{2}\left(q_{cm}^2+p_{cm}^2\right)+\frac{\hbar \omega_{r}}{2}\left(q_{r}^2+p_{r}^2\right)+
\frac{\hbar(\omega_2-\omega_1)G_0^1 G_0^2}{[G_0^1]^2+[G_0^2]^2} \left(q_{cm}q_{r}+p_{cm}p_{r}\right),
\end{equation}
where $\omega_{cm}=\left\{[G_0^1]^2\omega_1+[G_0^2]^2\omega_2\right\}/\left\{[G_0^1]^2+[G_0^2]^2\right\}$, and
$\omega_r=\left\{[G_0^1]^2\omega_2+[G_0^2]^2\omega_1\right\}/\left\{[G_0^1]^2+[G_0^2]^2\right\}$. This shows that at resonance,
$\omega_2=\omega_1$, the relative coordinate $q_r$ is decoupled from the center-of-mass and therefore also from the cavity mode. As a
consequence, it remains in its initial thermal state at the reservoir temperature and it is uncooled. Therefore, even though the center-of-mass
is cooled close to its ground state, the two mechanical modes, 1 and 2, are only negligibly cooled because their steady state energy is
determined by a weighted sum of the center-of-mass and relative mean energy. If instead $\omega_1 \neq \omega_2$, the relative motion is coupled
to the center-of-mass and the cooling of the latter by the cavity mode is able to partially cool also the relative motion. This ``sympathetic''
cooling is more efficient for larger coupling, i.e., for increasing $\omega_2-\omega_1$, provided that both modes are not too far from resonance
with the cavity.

\section{Entanglement properties of the steady state of the system}

From the results of Ref.~\cite{genes08} and \cite{prl07} one can see that in the optimal cooling regime for a single mechanical mode
$\Delta_{(1)} \simeq\omega_{1}$ one has also a significant stationary entanglement between the mechanical and the optical cavity mode, which is
also quite robust against temperature \cite{prl07}. It is therefore interesting to study the entanglement properties of the stationary state of
the tripartite system formed by the two close mechanical modes and the cavity mode. In particular, it is interesting to see if each mechanical
mode is entangled with the optical mode, as in the single mode case, and also if the common interaction with the cavity mode enables to
establish purely mechanical entanglement between the two vibrational modes.

The $6\times6$ steady state correlation matrix $\mathcal{V}$ defined by
Eq.~(\ref{CM}), particularized for the case of two mechanical modes, can be
written in terms of blocks of $2\times2$ matrices as
\begin{equation}
\mathcal{V}=%
\begin{pmatrix}
\mathcal{A}_{1} & \mathcal{C}_{12} & \mathcal{D}_{1}\\
\mathcal{C}_{12}^{\intercal} & \mathcal{A}_{2} & \mathcal{D}_{2}\\
\mathcal{D}_{1}^{\intercal} & \mathcal{D}_{2}^{\intercal} & \mathcal{B}%
\end{pmatrix}
. \label{CM2modes}%
\end{equation}
In order to quantify the bipartite entanglement of the Gaussian steady state of the three different instances of bipartite systems, we use the
logarithmic negativity \cite{logneg}, defined as
\begin{equation}
\mathcal{E}\left(  \mathcal{V}_{bip}\right)  =\max\{0,-\ln2\eta^{-}\left(
\mathcal{V}_{bip}\right)  \},
\end{equation}
where $\mathcal{V}_{bip}$ is a generic $4\times4$ correlation matrix
associated to the bipartite system of interest
\begin{equation}
\mathcal{V}_{bip}\equiv%
\begin{pmatrix}
\mathcal{A} & \mathcal{C}\\
\mathcal{C}^{\intercal} & \mathcal{B}%
\end{pmatrix}
,
\end{equation}
and $\eta^{-}\left(  \mathcal{V}_{bip}\right)  $ is given by \cite{logneg}
\begin{equation}
\eta^{-}\left(  \mathcal{V}_{bip}\right)  \equiv\frac{1}{\sqrt{2}}\left(
\Sigma\left(  \mathcal{V}_{bip}\right)  -\sqrt{\Sigma\left(  \mathcal{V}%
_{bip}\right)  ^{2}-4\det\mathcal{V}_{bip}}\right)  ^{1/2},
\end{equation}
with $\Sigma\left(  \mathcal{V}_{bip}\right)  \equiv\det\mathcal{A}
+\det\mathcal{B}-2\det\mathcal{C}$. The bipartite state is entangled if and
only if $\eta^{-}\left(  \mathcal{V}_{bip}\right)  <1/2$, which is equivalent
to the positive partial transpose (PPT) criterion, which is a necessary and
sufficient entanglement criterion in the case of bipartite Gaussian states
\cite{simon}.

\begin{figure}[h]
\centerline{\includegraphics[width=1.0\textwidth]{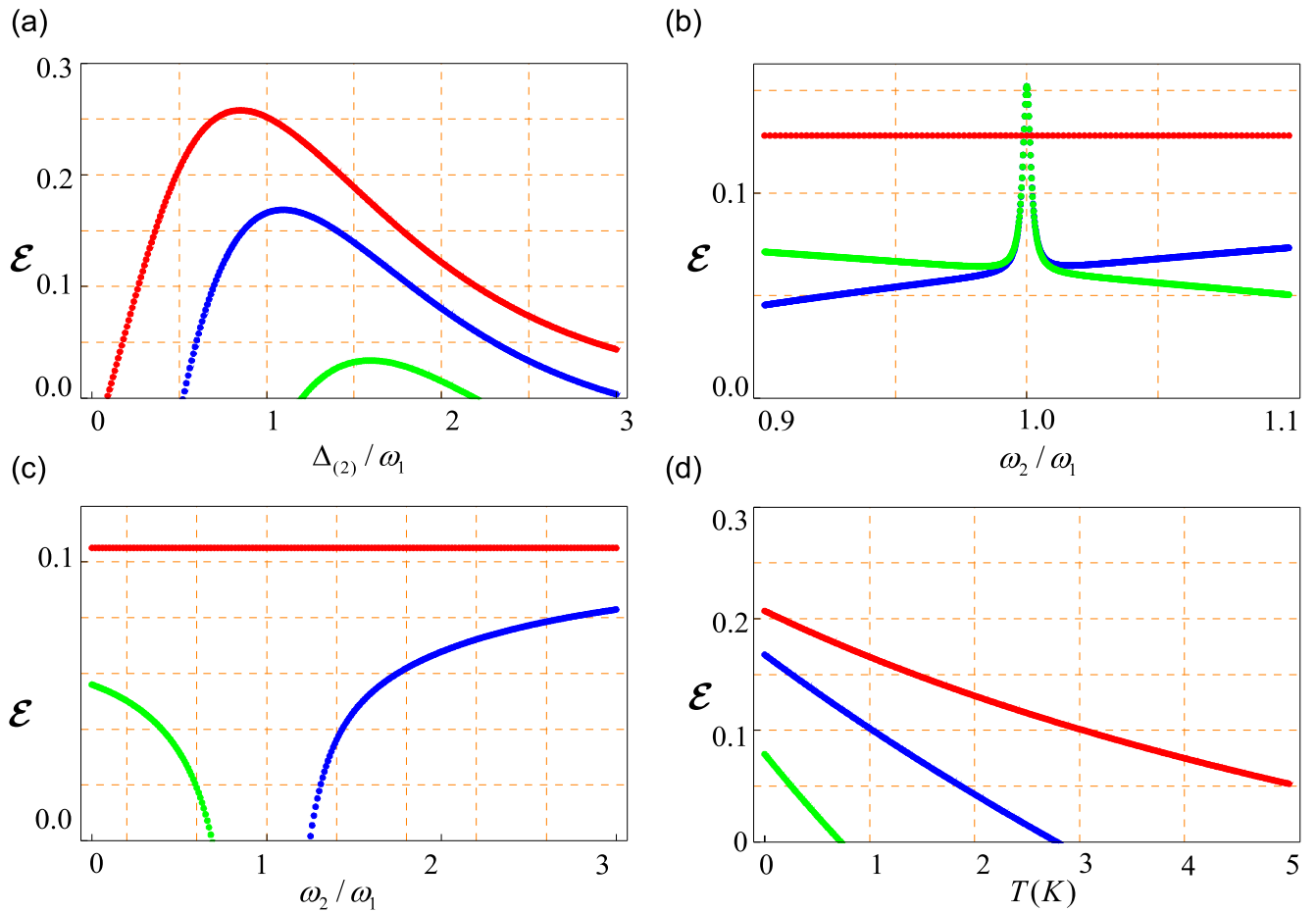}} \caption{Optomechanical entanglement. a) Mirror-field entanglement in the
absence of secondary mechanical modes (red line) is plotted versus normalized detuning $\Delta_{(2)}/\omega_{1}$ and compared with main
mode-field entanglement (blue) and secondary mode-field entanglement (green). The parameters are $\gamma_{1}/2\pi=\gamma_{2}/2\pi=100$ Hz,
$\omega_{1}/2\pi=10$ MHz, $\omega_{2}=1.5\times\omega_{1}$, $m_{1}=m_{2}=250$ ng, $T=0.4$ K, $\kappa=0.9\times\omega_{1}$ and
$G_{1}=\omega_{1}$. b) Enhancement of acousto-optical entanglement at zero temperature for close modes. The red line shows the value of the
negativity of the main mode-field entanglement in the absence of the secondary mode, while the blue and green curves show the behaviour of the
logarithmic negativity in the two-mode case when $\omega_{1}$ is fixed and $\omega_{2}$ is sweeped around $\omega_{1}$. In this case $G_1 \simeq
0.6 \omega_1$ and we have fixed $\Delta_2=\omega_1$. c) The enhancement shown in b) is lost as soon as the temperature increases. Here the
environment is at $T=0.4$ K and the secondary mode frequency is varied between $0.5\times\omega_{1}$ and $3\omega_{1}$. d) Temperature
robustness of entanglement in the collective case (blue and green lines for main and secondary mode respectively) compared to the independent
case (red line). Here $\omega_{2}=1.5\times\omega_{1}$ and we have chosen $\Delta_2=\omega_{2}$, such that a simultaneous entanglement regime is
obtained.}
\label{entanglement}
\end{figure}

To analyze the entanglement between one of the acoustic modes and the field, it suffices to eliminate the rows and columns that corresponds to
the other mirror mode from the matrix $\mathcal{V}$ of Eq.~(\ref{CM2modes}). We are left with a $4\times4$ matrix
\begin{equation}
\mathcal{V}_{f-m}\equiv%
\begin{pmatrix}
\mathcal{A}_{1,2} & \mathcal{D}_{1,2}\\
\mathcal{D}_{1,2}^{\intercal} & \mathcal{B}%
\end{pmatrix}
.
\end{equation}
We have performed a numerical analysis around the parameter region considered in Ref. \cite{prl07}, which is within reach of state-of-the-art
experiments (see the caption of Fig.~\ref{entanglement}) and for which, in the presence of a single mechanical mode, one has a significant,
stationary, optomechanical entanglement.

We find that, similar to what happens for cooling, the entanglement properties of the steady state of the system strongly depend upon the value
of the difference between the two mechanical resonance frequencies, $\delta \omega_{21}$. When the two modes are well separated, $\delta
\omega_{21}> \Gamma_i$, the presence of the second mode does not affect too much the main mode-cavity field entanglement. Moreover, also the
secondary mechanical mode is entangled with the cavity. This is illustrated in Fig.~\ref{entanglement}(a), where the optomechanical logarithmic
negativity in the single mechanical mode case (red line) is plotted versus the cavity detuning and compared with the corresponding curves for
the main mechanical mode (blue) and the secondary mode (green), when both modes are present. Optomechanical entanglement is more fragile than
cooling because in the two mode case it is always smaller than that with a single mechanical mode.

Figs.~\ref{entanglement}(b) and (c) show that the situation changes drastically when the two mechanical modes become very close in frequency,
$\omega_2/\omega_1 \simeq 1$. Both figures show the logarithmic negativity at a fixed detuning, versus the ratio $\omega_2/\omega_1$, at zero
temperature (b) and at $T=0.4$ K (c). We see that at zero temperature, the entanglement \emph{increases} around resonance, $\omega_2/\omega_1 =
1$. However, such an entanglement is very fragile with respect to temperature, and vanishes at $T=0.4$ K (c) for a wide interval around the
resonance condition. This behavior can be understood using the arguments of the preceding section. At mechanical resonance, the cavity mode is
strongly coupled, and entangled, to the center-of-mass, and uncoupled from the relative coordinate. Modes 1 and 2 are linear combinations of
$q_{cm}$ and $q_r$ and therefore their entanglement with the cavity mode is determined by both $q_{cm}$ and $q_r$. At $T=0$, both mode 1 and
mode 2 are entangled with the cavity mode thanks to the cavity-center-of-mass entanglement and because the quantum fluctuations of $q_r$ do not
significantly affect it. However, as soon as temperature is increased, the thermal fluctuations of the uncoupled coordinate $q_r$ kill the
entanglement of modes 1 and 2 with the cavity, even though cavity-center-of-mass entanglement is robust against temperature. Finally, the
robustness of these optomechanical entanglements against temperature is analyzed in Fig.~\ref{entanglement}(d). We see that even though never
comparable to the case of a single mechanical mode, provided that the frequencies of the two modes are sufficiently far apart, i.e., $\delta
\omega_{21} \gtrsim \Gamma_2$ (we have chosen $\omega_2=1.5\omega_1$), one has a regime in which the two modes are simultaneously entangled, and
this persists up to few Kelvins.

One can also check if the two mechanical modes, even though not directly interacting, can become entangled at the steady state thanks to the
common interaction with the cavity mode. Eliminating the entries in $\mathcal{V}$ that correspond to the cavity field, one is left with an all
mechanical correlation matrix
\begin{equation}
\mathcal{V}_{m-m}\equiv%
\begin{pmatrix}
\mathcal{A}_{1} & \mathcal{C}_{12}\\
\mathcal{C}_{12}^{\intercal} & \mathcal{A}_{2}%
\end{pmatrix}
.
\end{equation}
A numerical analysis of $\mathcal{E}\left(  \mathcal{V}_{m-m}\right)  $ in a parameter regime around the region of optimal cooling, i.e., that
of Figs.~\ref{cooling} and \ref{entanglement}, shows no entanglement between the two mechanical modes, even when they are both strongly
entangled to the same cavity field. Nonzero but extremely weak bipartite mechanical entanglement can be instead obtained in a regime where the
oscillators are heavily damped and the cavity finesse is very high ($\kappa\sim\gamma _{1,2}<\omega_{1,2}$). This is consistent with the results
of \cite{vitali07}, where bipartite entanglement between two different macroscopic oscillators is analyzed. In fact, the present system is
analogous to that of Ref.~\cite{vitali07}, with the center-of-mass $q_{cm}$ of the two modes here playing the same role of the relative
coordinate of Ref.~\cite{vitali07}. As already shown in Ref.~\cite{vitali07}, purely mechanical entanglement is very fragile with respect to
temperature and vanishes as soon as the occupancy of one of the modes is of the order of one.

\section{Classification of tripartite entanglement}

\begin{figure}[h]
\centerline{\includegraphics[width=1.0\textwidth]{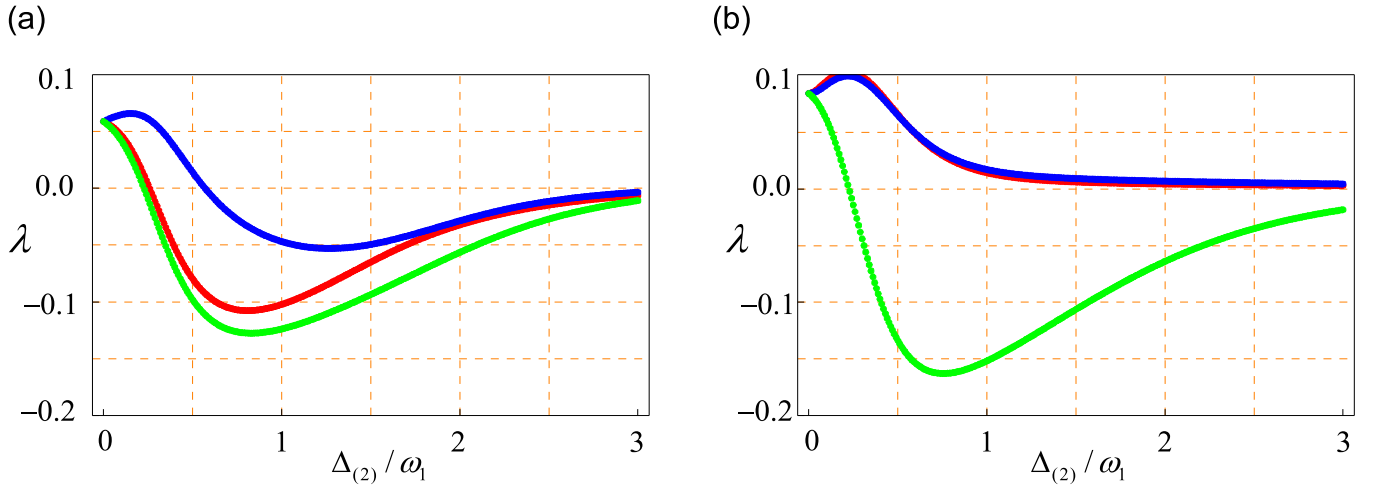}} \caption{Analysis of tripartite entanglement in the limit of well separated
mechanical modes $\omega_{2}=1.5\times\omega_{1}$ (a), and closely spaced modes $\omega_{2}=1.01\times\omega_{1}$ (b). The minimum eigenvalues
after partial transposition with respect to the main mode (red line), secondary mode (blue line) and field (green line) are plotted versus
normalized detuning $\Delta_{(2)}/\omega_{1}$ for a set of parameters $\gamma_{1}/2\pi=\gamma _{2}/2\pi=100$ Hz, $\omega_{1}/2\pi=10$ MHz,
$m_{1}=m_{2}=250$ ng, $T=0.4$ K, $L=0.5$ mm, $\kappa=0.5\times\omega_{1}$ and $G_{1}=0.7\times\omega_{1}$. For a large range of detunings, for
well separated modes the total system is in a fully inseparable tripartite entangled state (a) while for close frequencies
the resulting state is two-mode biseparable (b).}%
\label{trip}
\end{figure}

The tripartite system under study can have various forms of tripartite entanglement and it is therefore important to classify the entanglement
possessed by the steady state of the system. We determine the entanglement class of the system state by applying the results of
Ref.~\cite{giedke}, which has provided a necessary and sufficient criterion for the determination of the class in the case of tripartite CV
Gaussian states and which is directly computable. This classification criterion is mostly based on the nonpositive partial transposition (NPT)
criterion proved in \cite{entwerner}, which is necessary and sufficient for $ 1\times N$ bipartite CV Gaussian states. The NPT criterion of
\cite{entwerner} can be expressed in terms of the symplectic matrix
\begin{equation}
{\cal J} = \bigoplus_{i=1}^3 J_i, \;\;\;\;\;\; J_i=\left(
\begin{array}
[c]{cc}
0 & 1\\
-1 & 0
\end{array}
\right) \;\;\;\;\; {\rm i=1,2,3} \label{sympl}
\end{equation}
and of the partial transposition transformation $\Lambda_k$, acting on system $k$ only. Transposition is equivalent to time reversal and
therefore in phase space is equivalent to change the sign of the momentum operators. The NPT criterion states that a $1\times N$ CV Gaussian
state is separable if and only if the ``test matrix'' $\tilde{V}_{k}=\Lambda_{k}V\Lambda _{k}+i{\cal J}/2 \geq 0$. Therefore by evaluating the
sign of the eigenvalues of the three possible matrices $\tilde{V}_{k}$, and using the NPT criterion, one can discriminate between the various
entanglement classes. The results of this analysis are shown in Fig.~\ref{trip}. We find again that the properties of the steady state crucially
depend upon the comparison between the difference between the two mechanical frequencies $\delta\omega_{21}$ and the effective mechanical
damping $\Gamma_j$. When $\delta\omega_{21} > \Gamma_j$, the system is a fully tripartite state in a wide parameter region around the optimal
cooling regime studied here, because the minimum eigenvalues for each bipartition are always negative (see Fig.~\ref{trip}(a)). Instead, when
the two frequencies are very close to each other ($\delta\omega_{21} < \Gamma_j$), we see from Fig.~\ref{trip}(b) that two eigenvalues, those
corresponding to isolate a mechanical mode from the other two modes, are always positive, in a wide interval for the detunings. This means that
the steady state is of \emph{Class 3}, i.e., a two-mode biseparable state \cite{dur}, which is separable when the bipartite splits corresponding
to isolate one of the two mechanical modes are considered, but inseparable when the cavity mode is split from the rest. This can be understood
by recalling again that when $\omega_1=\omega_2$, the cavity mode is strongly coupled with the center-of-mass coordinate $q_{cm}$ and decoupled
from the relative coordinate $q_r$. The tripartite biseparable state manifests the fact that the cavity is entangled with the center of mass, of
the two oscillators. Entanglement is lost when one of the two mechanical modes is traced out and one restricts to bipartite entanglement only.

\section{Conclusions}

We have analyzed the effect of the radiation pressure of a cavity field mode on two vibrational modes of a mirror of Fabry-Perot cavity which
are near in frequency and both within the detection bandwidth. We have considered the effect of the second mechanical mode both on ground state
cooling via the back-action of the cavity mode, and on the optomechanical entanglement at the steady state. We have seen that the result
crucially depends upon the difference between the two mechanical resonance frequencies. If this difference is larger than the effective
bandwidth of the mechanical oscillators, given by the effective damping, the second mode not only does not affect cooling, but it is
simultaneously cooled together with the main mode. Under the same condition, each mode is entangled with the cavity mode and the steady state is
a fully tripartite entangled state. Instead when the two mechanical frequencies are very close to each other, cooling is destroyed by a
classical interference effect and both modes are uncooled. In this condition, each mechanical mode is entangled with the cavity mode at zero
temperature, but such an entanglement is extremely fragile with respect to temperature. Moreover, in the same regime, the steady state is a
two-mode biseparable state which is inseparable only when the cavity mode is split from the rest. In these parameter regimes, the mechanical
modes instead are never entangled; they can become entangled only for high finesse cavities, but the resulting entanglement vanishes at
extremely low temperatures.

\section{Acknowledgments}

This work was supported by the European Commission through the program QAP.

\end{document}